# Band offset and negative compressibility in graphene-$MoS_2$ heterostructures

*Stefano Larentis‡, John R. Tolsma‡ [a], Babak Fallahazad, David C. Dillen, Kyounghwan Kim, Allan H. MacDonald [a] and Emanuel Tutuc**

Microelectronics Research Center, Department of Electrical and Computer Engineering,

The University of Texas at Austin, Austin, TX 78758, USA

[a] Department of Physics, The University of Texas at Austin, Austin, Texas 78712, USA

**Abstract:** We use electron transport to characterize monolayer graphene - multilayer $MoS_2$ heterostructures. Our samples show ambipolar characteristics and conductivity saturation on the electron branch which signals the onset of $MoS_2$ conduction band population. Surprisingly, the carrier density in graphene decreases with gate bias once $MoS_2$ is populated, demonstrating negative compressibility in $MoS_2$. We are able to interpret our measurements quantitatively by accounting for disorder and using the random phase approximation (RPA) for the exchange and correlation energies of both Dirac and parabolic-band two-dimensional electron gases. This interpretation allows us to extract the energetic offset between the conduction band edge of $MoS_2$ and the Dirac point of graphene.

Transition metal dichalcogenides (TMDs) [1, 2] host a unique family of two-dimensional (2D) semiconductors with strong spin-orbit interactions, and coupled spin-valley degrees of freedom [3]. Because their in-plane effective masses are comparatively large, $m^* \approx 0.4 m_e$, where $m_e$ is the bare electron mass [4], electron-electron interaction effects are expected to be important in these semiconductors, even at relatively large electron densities. To date TMDs have been mostly employed as channel material in field effect transistors (FETs), in both top and back gated configurations, showing intrinsic mobility at room temperature up to 50 $cm^2/(V \cdot s)$ and on/off ratios larger than $10^5$ [5, 6, 7]. TMDs are also attractive for optoelectronic applications because of their high photoresponsivity [8], and thickness modulated band-gap [9]. Recently developed techniques make it possible to fabricate 2D material heterostructures [10], providing a pathway toward new devices including: graphene based tunneling transistors with hexagonal boron-nitride (hBN) [11] and $WS_2$ [12] tunneling barrier, memory elements based on graphene/insulator/$MoS_2$ stacks [13, 14], and graphene-$MoS_2$ heterostructures characterized by high photoresponsivity and gated persistent photoconductivity [15]. Nonetheless, the intrinsic electronic properties of $MoS_2$, and particularly the role of electron-electron interaction in these materials remain largely unexplored.

Here we report on a combined experimental and theoretical investigation of electron density partitioning in $MoS_2$-graphene heterostructures. Four point measurements of the heterostructure conductivity as a function of the back-gate bias show ambipolar characteristics, along with a clear saturation on the electron branch. Graphene layer carrier concentration measurements by magnetotransport reveal that the conductivity saturation is associated with carrier-population onset of the lower mobility $MoS_2$ layer. Experimental data from heterostructures with different $MoS_2$ thicknesses allows us to extract the band offset between the $MoS_2$ conduction band and the graphene charge neutrality point. Surprisingly, the carrier density in graphene decreases with increasing gate bias once electrons populate the $MoS_2$ layer, a finding associated with the negative compressibility of the $MoS_2$ electron system [16, 17]. To interpret our results, we solve the charge-partitioning problem using the thermodynamic equilibrium condition that the chemical potentials of $MoS_2$ and graphene electrons be equal. We find that the observation of decreasing graphene density above the critical voltage for $MoS_2$ occupation is a direct result of exchange and correlation energy contributions to the many-body chemical potential of $MoS_2$. We also introduce a simple model to account for the role of disorder in reducing the magnitude of interaction effects at the lowest carrier densities.

The graphene-$MoS_2$ heterostructures studied in this paper consist of monolayer graphene transferred onto a multilayered $MoS_2$ flake mechanically exfoliated onto a $SiO_2$/Si substrate, which serves as back-gate (Fig. 1a). To fabricate these samples we first exfoliate commercially available $MoS_2$ crystals (SPI Inc., 2D semiconductors) onto a 285 nm thick $SiO_2$ dielectric, thermally grown on a highly doped Si substrate. The exfoliated flakes are annealed in ultra-high vacuum (UHV) at 350°C to remove tape residues. Flake topography is then verified with atomic force microscopy (AFM) (Fig. 1b). Graphene monolayers are exfoliated from natural graphite onto a Si substrate coated with a polymethyl methacrylate (PMMA)/polyvinyl alcohol (PVA) stack. The PVA interlayer is then dissolved with deionized water separating the Si substrate from the PMMA membrane, which is moved onto a custom glass mask. Using a mask-aligner with a custom made heated stage, the membrane is aligned and brought into contact with the target $MoS_2$ flake [18]. Figure 1c shows the AFM micrograph of a monolayer graphene flake transferred onto $MoS_2$ after PMMA removal and UHV anneal. The transferred graphene displays some ripples, which are partially removed during the annealing process [19]. Electron beam lithography (EBL), $O_2$ plasma etching and UHV anneal are used to define a graphene Hall bar in a ripple-free region as shown in Fig. 1d. A second EBL step, followed by e-beam evaporation of Ni/Au and lift off is used to define the metal contacts. Figure 1e shows an optical micrograph of a completed device. Figure 1f shows Raman spectra of device regions with and without graphene, obtained using 532 nm excitation wavelength. The data exhibit the $E^1_{2g}$ and $A_{1g}$ peaks characteristic of $MoS_2$, as well as the G and 2D peaks for graphene. The fabricated samples were studied by magnetotransport at temperatures ($T$) down to 1.4 K and perpendicular magnetic fields ($B$) up to 14 T, using low-current, low-frequency lock-in techniques.

Figure 2a shows the heterostructure conductivity ($\sigma$) as a function of the back-gate bias ($V_{\mathrm{BG}}$) at temperatures ranging between 4.5 K and 295 K. The conductivity reaches a minimum at a gate bias value ($V_D$), which we identify with charge neutrality in the graphene sheet. For $V_{BG} < V_D$ ($V_{BG} > V_D$) $\sigma$ decreases (increases) with increasing $V_{BG}$, indicating holes (electrons) populate the heterostructure. While this ambipolar behavior is characteristic of graphene, $\sigma$ vs $V_{BG}$ data displays particle-hole asymmetry characterized by clear saturation at a positive threshold gate bias ($V_T$) which becomes more sharply defined upon reducing the temperature. We interpret these data using the band structure diagrams of Figs. 2b-2d. For $V_{BG} < V_D$ gate-induced carriers are added to the graphene valence band (Fig. 2b). For $V_D < V_{BG} < V_T$ the carriers are added to the graphene conduction band (Fig. 2c). For $V_{BG} < V_T$ the chemical potential in graphene is lower than the $MoS_2$ conduction band edge. At a sufficiently large gate bias, the increase in the chemical potential of electrons in graphene coupled with the electric field induced band bending brings the bottom of the conduction band of $MoS_2$ into alignment with the graphene chemical potential. For $V_{BG} > V_T$ carriers are added to the $MoS_2$ conduction band (Fig. 2d). The total carrier density ($n_{Tot}$) summed over graphene ($n_G$), and $MoS_2$ ($n_{MoS2}$) systems satisfies:

$$n_{Tot} = n_G + n_{MoS2} = \frac{C_{ox}}{e}(V_{BG} - V_D) \qquad (1)$$

where $C_{ox}$ is the back-gate oxide capacitance, and $e$ is the electron charge. Because the separation between the back gate and the heterostructure is much larger than the typical electronic screening lengths of either $MoS_2$ or graphene, quantum capacitance contributions to $C_{ox}$ are negligible. Indeed, the graphene quantum capacitance at a density of $10^{11}$ $cm^{-2}$ is approximately two orders of magnitude larger than the 285 nm thick $SiO_2$ dielectric capacitance, and can be neglected as a series contribution. We attribute the conductivity saturation beyond the threshold voltage to the lower $MoS_2$ mobility compared to graphene. Indeed, the graphene field-effect mobility extracted from Fig. 2a data for $V_{BG} < V_T$ is 8000 $cm^2/(V \cdot s)$, while typical electron mobilities we measured in separate four-terminal $MoS_2$ devices range between 10-50 $cm^2/(V \cdot s)$ at room temperature, increasing up to 1000 $cm^2/(V \cdot s)$ at 4 K [6, 20, 21]. These findings are consistent with previous measurements of the $MoS_2$ conductivity that range between 30 μS at room temperature to 300 μS at 4 K in the ON state [6, 7], not large enough to provide significant parallel conduction. We remark that the mobility of electrons in graphene supported by $MoS_2$ is significantly lower than that of graphene supported by hBN [18], even though the heterostructure fabrication techniques are very similar.

Figure 3a shows the four point longitudinal resistance ($R_{xx}$) measured as a function of $B$ at $T$ = 1.4 K, and at different $V_{BG}$ values. The gate bias value at neutrality in this device is $V_D$ = -15 V. $R_{xx}$ data display Shubnikov-de Haas (SdH) oscillations for $B$ fields as low as 2 T, and follow a quantum Hall state sequence (QHS) with filling factors $\nu$ = ±2, 6, 10, 14, 18 that can be attributed to monolayer graphene [22]. Figure 3b shows an example of $R_{xx}$ and $R_{xy}$ vs. $B$ data measured at $V_{BG}$ = 20 V, with the corresponding QHS filling factors $\nu$ = + 6, 10, 14 marked. The carrier density in graphene ($n_G$) at a fixed $V_{BG}$, is extracted from the $R_{xx}$ oscillations minima using:

$$n_G = \frac{eB_\upsilon}{h}\upsilon \qquad (2)$$

where $B_\nu$ is the magnetic field corresponding to the QHS at filling factor $\nu$, marked in Fig. 3a, and $h$ is the Plank constant. Because separate magnetotransport measurements on $MoS_2$ layers do not show SdH oscillations, we associate the $R_{xx}$ minima exclusively with QHSs in graphene. The $MoS_2$ carrier density ($n_{MoS2}$) can be inferred using the experimentally determined $n_G$ values and the total density calculated using Eq. (1). Figure 3c shows $n_G$ and $n_{MoS2}$ vs $V_{BG}$, along with the $\sigma$ vs $V_{BG}$ data measured at $B$ = 0 T. Two main findings are apparent from Fig. 3c. First, the $n_G$ vs $V_{BG}$ data shows a clear feature at $V_T$, concomitant with the saturation of $\sigma$ vs. $V_{BG}$ data. Secondly, and perhaps most surprisingly, the carrier density in graphene *decreases* with increasing gate bias for $V_{BG} > V_T$, a finding which can be explained by the influence of exchange and correlation energy on the many-body chemical potential of $MoS_2$. In the following, we further analyze these two findings.

Fig. 4a shows the heterostructure band diagram at flat-band: $V_{BG} = V_D$, and $n_G$ = 0. We introduce here the band offset ($\Delta E_C$) as the energy separation between the graphene charge neutrality point and the $MoS_2$ conduction band edge. In order for the $MoS_2$ to be populated with electrons, the $MoS_2$ conduction band edge has to be brought into alignment with the chemical potential in graphene. In a gated structure, this can be accomplished thanks to the increase in graphene chemical potential at $V_{BG} > V_D$, along with the electrostatic band bending of the $MoS_2$ conduction band. Equilibrium is achieved between graphene and $MoS_2$ systems when their chemical potentials, including electrostatic band bending and exchange-correlation contributions, are equal:

$$\mu_G(n_G) = \mu_{MoS2}(n_{MoS2}) + \Delta E_C - \frac{e^2 t_{MoS2}}{\epsilon_{MoS2}^{\parallel} \epsilon_0} n_G \qquad (3)$$

Here, $\mu_G(n_G)$ and $\mu_{MoS2}(n_{MoS2})$ are the graphene and $MoS_2$ chemical potentials including exchange-correlation contributions at carriers densities $n_G$ and $n_{MoS2}$, measured from the neutrality point and $MoS_2$ conduction band edge, respectively; $t_{MoS2}$ and $\epsilon_{MoS2}^{\parallel}$ are respectively the $MoS_2$ thickness and relative dielectric constant parallel to the $c$-axis; $\epsilon_0$ is the vacuum dielectric permittivity. The last term in Eq.(3) accounts for the electrostatic band bending of the $MoS_2$ conduction band edge, assuming that $t_{MoS2}$ is the effective interlayer separation between graphene and the occupied states in our few layer $MoS_2$ samples, as explained below.

Interlayer coupling in transition metal dichalcogenides is weak, as indicated by the small energy band width (≈ 30-40 meV) of bulk $MoS_2$ conduction bands along the high symmetry direction K-H [23, 24]. At the onset of $MoS_2$ population, the transverse electric field induced by the graphene carrier density $n_G = 0.8 \times 10^{12}$ cm$^{-2}$ creates a potential drop across neighboring $MoS_2$ layers of 0.025 eV, comparable to the interlayer hopping energy scale. We thus choose to model our results by approximating the $MoS_2$ band structure by a set of uncoupled monolayers. Using the two dimensional $MoS_2$ density of states (DOS) of $3.6 \times 10^{14}$ eV$^{-1}\cdot$cm$^{-2}$ yields a minimum density for the second $MoS_2$ band (layer) to be occupied of $n_{MoS2} = 9 \times 10^{12}$ cm$^{-2}$; for smaller densities, electrons occupy only the layer with lowest electrical potential (i.e. the layer furthest from graphene and closest to the gate). The opposite limit of this approximation is to neglect the potential drop across neighboring layers of $MoS_2$ and model the carriers as occupying the lowest sub-band of a few-layer system. The principle change to our analysis would be to reduce the effective value of the graphene-$MoS_2$ separation. For example, the sub-band energy splitting near the conduction band minimum in six-layer $MoS_2$ is $\approx$ 0.1 eV [25], a value corresponding to a $MoS_2$ carrier density of nearly $4 \times 10^{13}$ $cm^{-2}$ necessary to occupy the second subband. These considerations validate the approximation we use, namely treating $MoS_2$ as a single band two-dimensional electron gas.

Figure 4b shows $\sigma$ vs $V_{BG}$ data for graphene/$MoS_2$ heterostructures with different $MoS_2$ thicknesses, measured at temperatures between 1.4 K and 10 K. The $MoS_2$ population threshold is marked on each trace. Experimental $V_T$ - $V_D$ values as a function of $MoS_2$ sample thickness $t_{MoS2}$ are summarized in Fig. 4c. Figure 4c data can be fitted using $n_G = \frac{C_{ox}}{e}(V_T - V_D)$ and $\Delta E_C$ can be extracted using Eq. (3) with $\mu_{MoS2}$ set to zero. The fit to the experimental data, assuming the experimental $C_{ox}$ = 12.1 nF/$cm^2$ value, and a theoretical $\epsilon^{\parallel}_{MoS2}$ = 4 [26], yields $\Delta E_C$ = 0.29 eV. We note here that since the $MoS_2$ band gap depends on thickness as the monolayer limit is approached, the extracted $\Delta E_C$ value is most accurate for thicker samples. It is also informative to compare the experimentally determined graphene-$MoS_2$ band offset with graphene work function (WF) and $MoS_2$ electron affinity ($\chi_{MoS2}$) differences. The experimental value for the graphene WF is 4.57 ± 0.05 eV [27, 28]. Experimental studies of electron affinity in bulk $MoS_2$ report $\chi_{MoS2} \approx$ 4.2 eV [29]. While data in the limit of few layer $MoS_2$ films are presently lacking, theoretical studies report a constant $\chi_{MoS2} \approx$ 4.4 eV for $MoS_2$ thicker than 4 layers, a range which corresponds to the $MoS_2$ thickness probed in our study [30]. The conduction band offset extracted using these values ranges between 0.17-0.37 eV, a range consistent with $\Delta E_C = 0.29$ eV experimentally determined in this study.

We now describe how to account for our observations quantitatively by including many-body exchange and correlation, as well as disorder. We approximate the only occupied $MoS_2$ conduction sub-band by a two-valley two-dimensional electron gas (2V2DEG) with effective mass $m^* = 0.43m_e$ [4]. At low density (Fermi wave-vector, $k_F$, much smaller than the inverse lattice constant) inter-valley electron-electron scattering is strongly suppressed relative to intra-valley scattering by the long range of the Coulomb potential. The valley degree of freedom then enters our expressions for the interaction energy only as an effective degeneracy factor, $g_V = 2$. The exchange energy per electron ($\varepsilon_{ex}$) of the 2V2DEG can be evaluated analytically and is given by:

$$\varepsilon_{ex} = -\frac{16}{3\,\pi\,(g_V g_S)^{\frac{1}{2}}}\left(\frac{Ry^*}{r_S}\right) \qquad (4)$$

where $g_s = 2$ is the spin degeneracy, and $Ry^* = 184$ meV is the Rydberg energy reduced by a factor of $m^*/\kappa^2$, where the form of the 2V2DEG dielectric constant $\kappa = \left(\sqrt{\epsilon_{MoS2}^{\parallel}\epsilon_{MoS2}^{\perp}} + \sqrt{\epsilon_{SiO2}^{\parallel}\epsilon_{SiO2}^{\perp}}\right)/2$ reflects the anisotropy of the dielectric material surrounding the occupied $MoS_2$ band (layer). The product of the $MoS_2$ Fermi surface diameter ($2k_{\mathrm{F}}$) and the $MoS_2$ layer thickness ($t_{MoS2}$) can be used to assess both the relevance of the dielectric screening in the vacuum region, as well as correlations between $MoS_2$ and graphene electrons [31]. For carrier densities large enough that the influence of disorder on the $MoS_2$ chemical potential is minimal, we find $2k_{\mathrm{F}}t_{MoS2} \approx 3-6$, implying a negligible role for this remote part of the $MoS_2$ 2DEG environment. We use a theoretical bulk value of $\epsilon_{MoS2}^{\perp} = 13.5$ [26] for the dielectric constant perpendicular to the $c$-axis, and as noted above we employ $\epsilon_{MoS2}^{\parallel} = 4$ [26]. With $\epsilon_{SiO2}^{\parallel} = \epsilon_{SiO2}^{\perp} = 3.9$ we find $\kappa = 5.6$. The parameter $r_s = 1/\left(a_B^* \sqrt{\pi n}\right)$, where $n$ is the carrier density and $a_B^* = \left(\frac{\kappa}{m^*}\right) a_B = 6.9$ Å the effective Bohr radius, is proportional to the ratio of interaction to kinetic energy in continuum electron gas models [17], and reaches relatively large values of $r_s = 8$ - $26$ in $MoS_2$ at moderately high electron densities in the range $n_{MoS2} = 10^{11}$ -$10^{12}$ $cm^{-2}$. By comparison, similar $r_s$ values in GaAs 2D electron systems are reached at densities nearly two orders of magnitude lower.

To evaluate the correlation energy per particle we follow the common procedure of combining coupling-constant integration with the fluctuation-dissipation relationship between the density structure factor and the RPA density response function [17]. The correlation energy per electron ($\varepsilon_c$) can be isolated and written as an integration over the dimensionless wavevector, $q$, frequency, $\omega$, and the imaginary axis Lindhard function, $\chi(q, \mathbb{i}\omega)$ [17]:

$$\varepsilon_c = \frac{4}{g_V^2\, g_S^2}\left(\frac{Ry^*}{\pi\, r_S^2}\right) \int q\, dq \int d\omega \left( r_S \frac{(g_V g_S)^{\frac{3}{2}}}{2\, q} \chi(q, \mathbb{i}\omega) + ln\left(1 - r_S \frac{(g_V g_S)^{\frac{3}{2}}}{2\, q} \chi(q, \mathbb{i}\omega)\right)\right) \qquad (5)$$

Both the exchange and correlation energies of the 2V2DEG are negative, and reduce the total energy per particle. The magnitude of the correlation energy per electron in 2V2DEGs is slightly enhanced compared to single valley two-dimensional electron systems, but the reduction in exchange energy magnitude more than compensates, and the total interaction energy is slightly reduced in magnitude for all values of $r_s$. Interestingly, recent quantum Monte Carlo calculations [32] have found that the disorder free 2V2DEG remains paramagnetic down to the extreme low densities necessary for Wigner crystallization ($r_s = 45$), never undergoing a Bloch ferromagnetic transition [17].

We have used the same procedure to calculate the RPA ground state energy per particle of doped graphene. This calculation evaluates expressions given in reference [33] using a dielectric constant of $\kappa_G = \left(1 + \left(\epsilon^{\parallel}_{MoS2}\epsilon^{\perp}_{MoS2}\right)^{1/2}\right)/2 = 4.2$. Correlation and exchange in graphene act oppositely and the overall interaction effect is thus smaller in relative terms. Our calculations shows that near the graphene density saturation, its many-body chemical potential is accurately approximated by the non-interacting expression, $\mu_G(n_G) = \hbar v_F \sqrt{\pi\, n_G}$, where $v_F$ is the Fermi velocity in graphene, with a numerical value larger by approximately 20% with respect to the bare value $v_{F0} = 1 \times 10^8$ cm/s, in close agreement with recent experimental results [34].

The many-body chemical potential for both graphene and $MoS_2$ is calculated from the ground state energy by numerical differentiation using $\mu(n) = \partial(n\,\varepsilon(n))/\partial n$, where $\varepsilon(n) = \varepsilon_0(n) + \varepsilon_{ex}(n) + \varepsilon_c(n)$ and $\varepsilon_0(n)$ is the non-interacting energy per electron. To illustrate the relative contribution of exchange and correlation to the chemical potential in $MoS_2$, in Fig. 5a we plot $\mu_{MoS2}$ vs. $n_{MoS2}$, with and without the interaction contributions. In the entire range of densities experimentally accessible, the interaction contribution to the chemical potential is negative and much larger than the kinetic energy contribution. Note that the RHS of Eqn. (3), the equilibrium charge partition equation, depends on $\mu_{MoS2}$, but also on the conduction band offset from the Dirac point and band bending from the electric field of the gate. In Fig. 5b-c we plot the LHS and RHS of Eq. (3), as a function of $n_G$, at two $n_{Tot}$ values (i.e. fixed gate voltage), for a graphene-$MoS_2$ heterostructure with $t_{MoS2}$ = 4.2 nm, similar to Fig. 3 data. The $n_{Tot}$ value in Fig. 5b corresponds to the threshold of $MoS_2$ population, in good agreement with the experimental data in Fig. 3c. We remark here that in a limited $n_{Tot}$ range, Eq. (3) has two solutions for $n_G$ and $n_{MoS2}$, suggesting a possible charge bistability. We find that the solution with the smaller $n_{MoS2}$ value occurs at a maximum in energy per volume, whereas the solution at larger $n_{MoS2}$ occurs at a minimum and is therefore energetically favorable.

To better compare theory and experiment, in Fig. 6 we plot $n_G$ vs. $V_{BG}$ for the same graphene-$MoS_2$ heterostructure considered in Fig. 5 with $t_{MoS2}$ = 4.2 nm. The lines represent calculated values, while the symbols are experimental data from Fig. 3c. To illustrate the role of exchange and correlation, Fig. 6 includes the calculated $n_G$ vs $V_{BG}$ neglecting (dashed red) and including (dark blue) the electron-electron interaction contributions to $\mu_{MoS2}$ and $\mu_G$. The reduction in graphene density above the threshold for $MoS_2$ occupation is a direct result of these interactions, and is well captured by our RPA theory for the many-body chemical potentials of $MoS_2$ and graphene. Theoretical results indicate a small discontinuous change in the layer carrier density (~ $3 \times 10^{10}$ cm$^{-2}$) at the onset of $MoS_2$ occupation. The absence of a jump in the experimental data can be explained by considering the impact of disorder in $MoS_2$. Similar observations were made in double layer GaAs/AlGaAs quantum wells [16], where at carrier densities below $10^{10}$ cm$^{-2}$, disorder obscures the otherwise expected divergence in compressibility.

Using a typical $MoS_2$ low temperature mobility value of 500 cm$^2$/(V·s), which corresponds to a scattering time $\tau = 1.2 \times 10^{-13}$ s, we obtain a disorder energy scale $\Gamma_{dis} = \hbar/\tau = 5.4$ meV. When the Fermi energy in $MoS_2$ is of this magnitude or less, we can expect disorder to play a significant thermodynamic role. Because of its large band mass, the $MoS_2$ Fermi energy is lower than 5.4 meV for $n_{MoS2} \lesssim 2 \times 10^{12}$ cm$^{-2}$, a sizeable fraction of the density range experimentally accessible. To quantitatively account for the influence of disorder in $MoS_2$, we introduce a simple model for the $MoS_2$ chemical potential based on the following phenomenological approximation for the DOS of disordered $MoS_2$,

$$g_{dis}(E) = \frac{g_0}{2}\left(Tanh\left(\frac{E - \Delta E_c}{\Gamma_{dis}}\right) + 1\right) \qquad (6)$$

For electron energies well above the band edge, the DOS is $g_0 = 2m^*/(\pi \hbar^2)$. Equation (6) captures disorder induced smearing of the DOS jump that occurs in the absence of disorder [35]. The single-particle contribution to the chemical potential corresponding to Eq. (6) is

$$\mu_{dis}(n_{MoS2}) = \frac{\Gamma_{dis}}{2}\ln\left(exp\left(\frac{2\, n_{MoS2}}{g_0\, \Gamma_{dis}}\right) - 1\right) \qquad (7)$$

Because $\mu_{dis}(n_{MoS2})$ increases rapidly with $n_{MoS2}$ at low carrier densities, disorder counteracts in part the otherwise dominant influence of exchange and correlation on the chemical potential. We solve the equilibrium problem replacing the kinetic energy contribution to the chemical potential by Eq. 7. In Fig. 6 we compare calculations including and neglecting disorder. The $MoS_2$ band edge offset is fit by requiring $n_G = 8.6 \times 10^{11}$ cm$^{-2}$ at $V_{BG} = 40$ V. At this large gate voltage disorder in the $MoS_2$ plays a relatively weaker role. Fitting at $\Gamma_{dis} = 5.4$ meV, corresponding to a $MoS_2$ mobility of 500 cm$^2$/(V·s) yields the same band offset $\Delta E_C = 0.41$ eV obtained above using the disorder-free theory. Fitting with a larger $\Gamma_{dis} = 26.9$ meV, corresponding to a $MoS_2$ mobility of 100 cm$^2$/(V·s) give a slightly different band offset value of $\Delta E_C = 0.42$ eV, demonstrating that disorder effects do not limit our ability to extract the offset.

When disorder is included, the discontinuous drop in graphene density at the onset of $MoS_2$ occupation no longer occurs, and the maximum graphene density achieved prior to $MoS_2$ occupation is reduced, both features improving agreement with experiment.

Lastly, we comment on the relevance of these results for potential device applications. The $MoS_2$ negative compressibility translates into a negative quantum capacitance of $MoS_2$ electrons in a FET. While quantum effects, such as quantum capacitance and inversion layer thickness typically reduce the FET gate capacitance with respect to the dielectric capacitance, the negative $MoS_2$ quantum capacitance can enable FETs with a gate capacitance *larger* than the dielectric capacitance.

In summary, we investigate the electrical properties and the carrier distribution in a graphene-$MoS_2$ heterostructure. The conductance-density dependence reveals a marked saturation on the electron branch, associated with the onset of $MoS_2$ conduction band population. This observation allows the graphene-$MoS_2$ band offset to be extracted from the data. Magnetotransport measurements show a surprising decrease of the graphene electron density with gate bias beyond the $MoS_2$ population threshold, a finding that highlights the negative compressibility of the $MoS_2$ electron system. Using the random phase approximation for the exchange-correlation contribution to the chemical potentials of $MoS_2$ and graphene, we solve for the density distribution as a function of back gate voltage (i.e. total density). The decrease in graphene density at large gate voltage is due entirely to interaction effects. We are able to account quantitatively for features at the onset of $MoS_2$ population and refine our band-offset estimate by using a phenomenological model for the density of states in disordered $MoS_2$. Theoretical calculations are in good agreement with experiment, and demonstrate that the $MoS_2$ electron gas is strongly correlated.

FIGURES

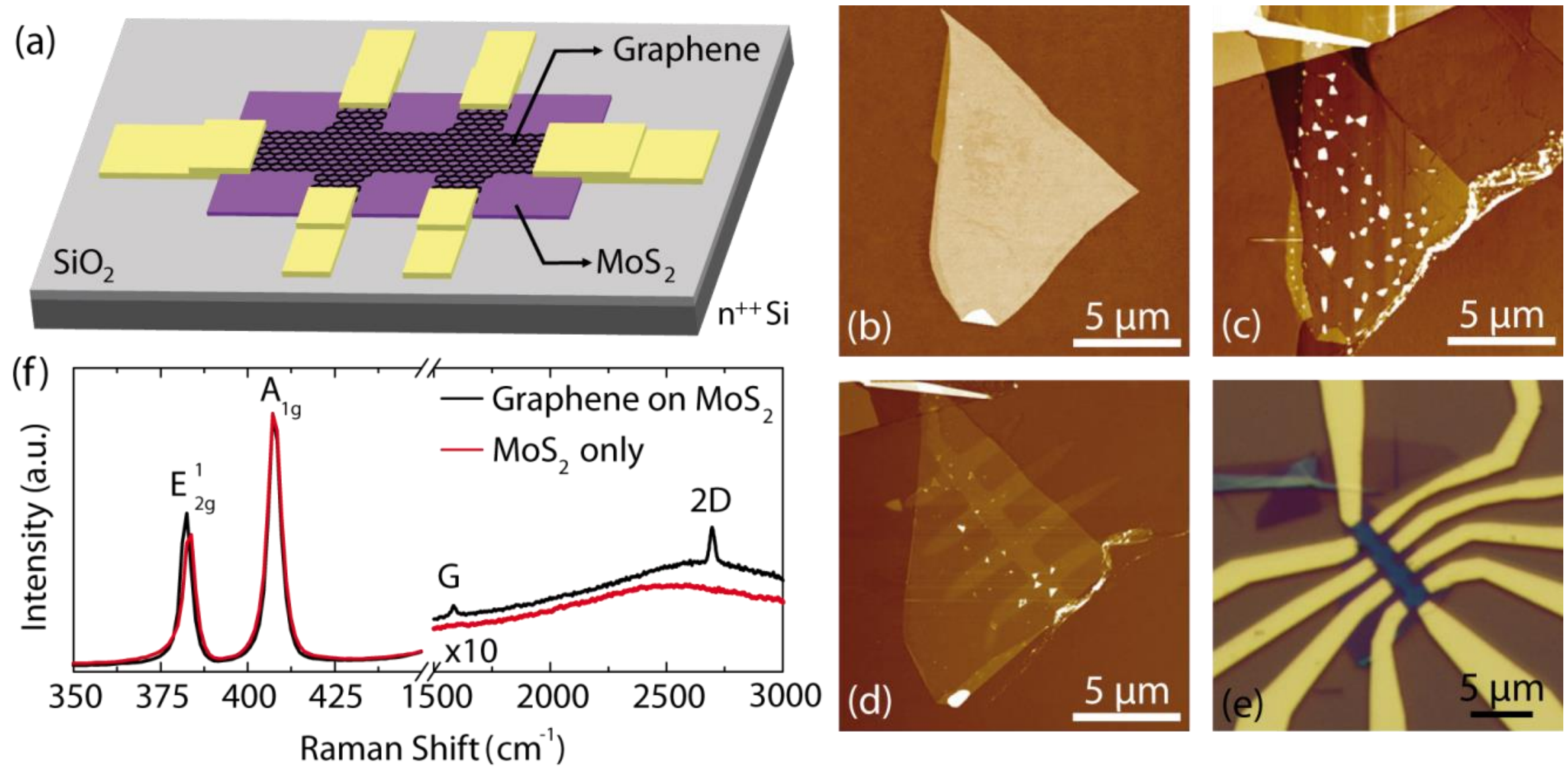

**Figure 1.** (a) Schematic representation of the graphene-$MoS_2$ heterostructure. (b) AFM topography of a $MoS_2$ flake exfoliated on $SiO_2$/Si substrate, after UHV anneal at 350° C. (c) AFM topography of monolayer graphene transferred on the same flake as in (b), after UHV anneal. Note the presence of ripples on the graphene surface, even after UHV anneal. (d) AFM topography of the graphene-$MoS_2$ heterostructure after the graphene Hall bar is patterned and the device annealed. (e) Optical micrograph of the completed device with Ni/Au contacts. (f) Raman spectra of device regions with and without graphene, obtained using 532 nm excitation wavelength.

.

**Figure 2.** (a) $\sigma$ vs. $V_{BG}$ measured at different $T$ ranging from 4.5 K to 295 K. The electron branch shows a clear saturation of σ for $V_{BG} > V_T$. The different shaded areas correspond to the band diagrams in (b), (c) and (d). (b), (c) Band diagram of the heterostructure for $V_{BG} < V_T$ ($V_D < V_{BG} < V_T$) when holes (electrons) are induced in the graphene layer. (d) Band diagram for $V_{BG} > V_T$ when electrons are induced in the $MoS_2$ conduction band.

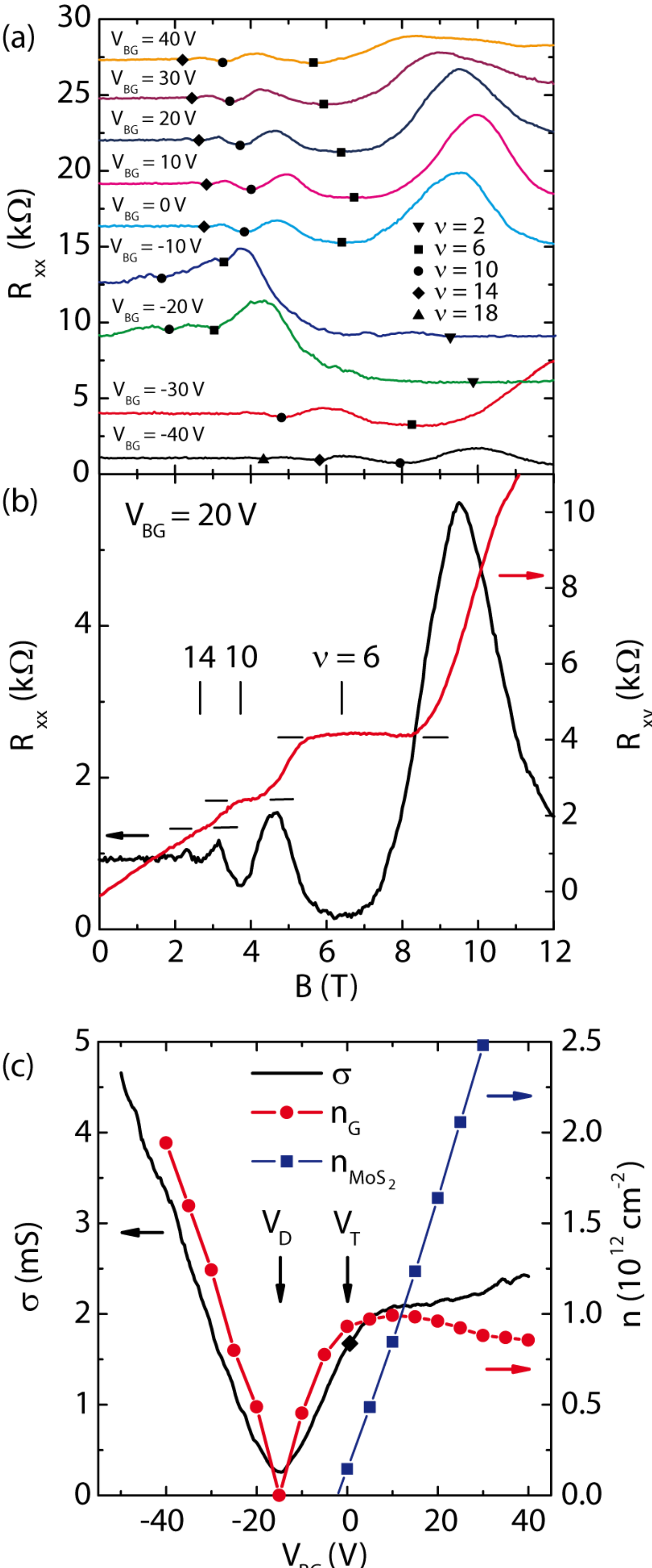

**Figure 3.** (a) $R_{xx}$ vs $B$ measured for different $V_{BG}$ (solid lines). The symbols mark the $R_{xx}$ oscillation minima position and the corresponding ν. The different traces are displaced vertically for clarity. (b) $R_{xx}$ and $R_{xy}$ vs. $B$ measured at $V_{BG}$ = 20 V. The QHSs corresponding filling factors are marked. (c) $\sigma$ vs. $V_{BG}$ at $B$ = 0 T (black line, left axis), $n_G$ vs $V_{BG}$ (red symbols, right axis) extracted from the SdH oscillations using panel (a) data, and $n_{MoS2}$ vs $V_{BG}$ (blue symbols, right axis) obtained as the difference between the total density and $n_G$. $\sigma$ and $n_G$ saturate for $V_{BG} > V_T$ , as the $MoS_2$ becomes populated. The $n_G$ decrease for $V_{BG} > V_T$ shows evidence of negative compressibility in the $MoS_2$ electron gas.

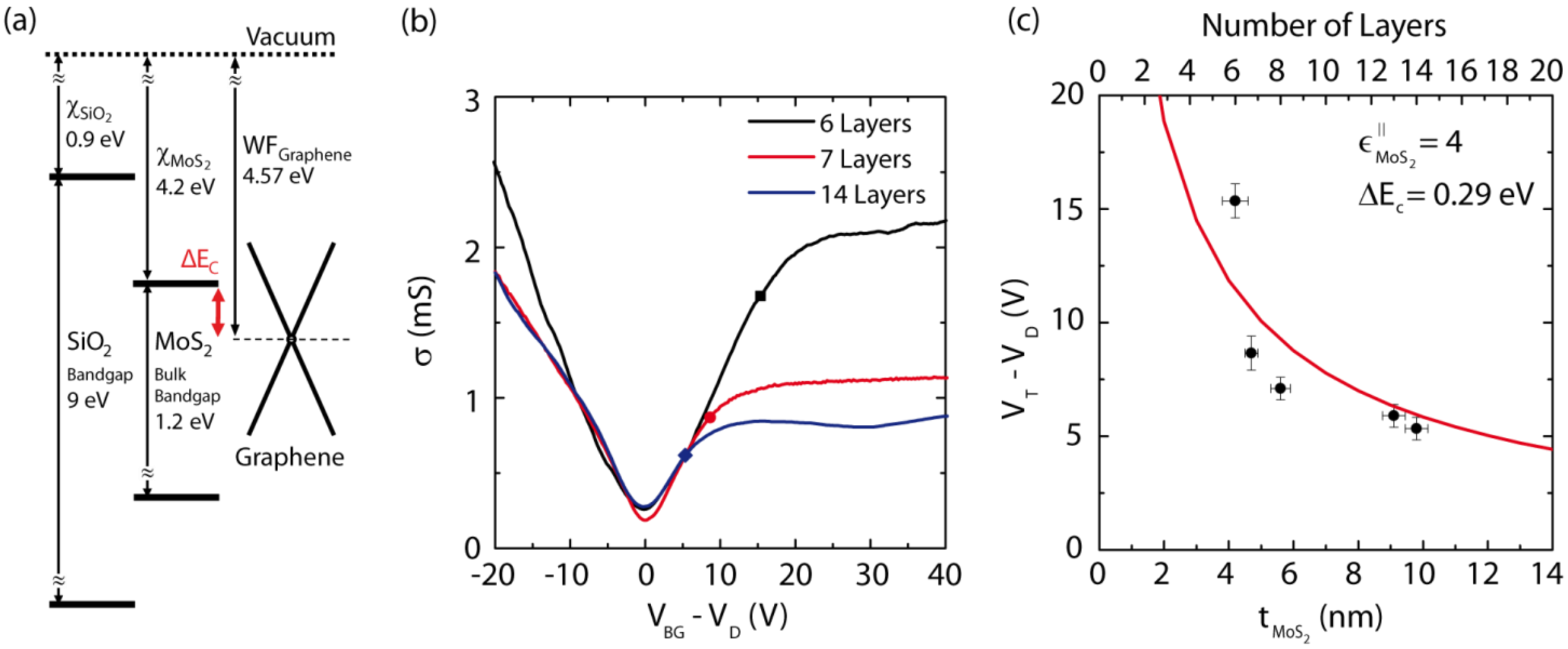

**Figure 4.** (a) Band diagram at flat band ($V_D$ = $V_{BG}$, $n_G$ = 0) in the graphene-$MoS_2$-$SiO_2$ heterostructure, constructed using data from [27, 28, 29]. (b) $\sigma$ vs $V_{BG}$ for different $t_{MoS2}$, namely 6, 7 and 14 layers (solid lines), measured at $T$'s between 1.4 - 10 K. The solid symbols mark the $V_T$ value on each trace. (c) $V_T$ - $V_D$ vs $t_{MoS2}$ (solid symbols), showing a $V_T$ - $V_D$ increase for decreasing $t_{MoS2}$. The fit (red line) to the experimental data using Eq. (3) yields $\Delta E_C$ = 0.29 eV, for $\epsilon^{\parallel}_{MoS2} = 4$.

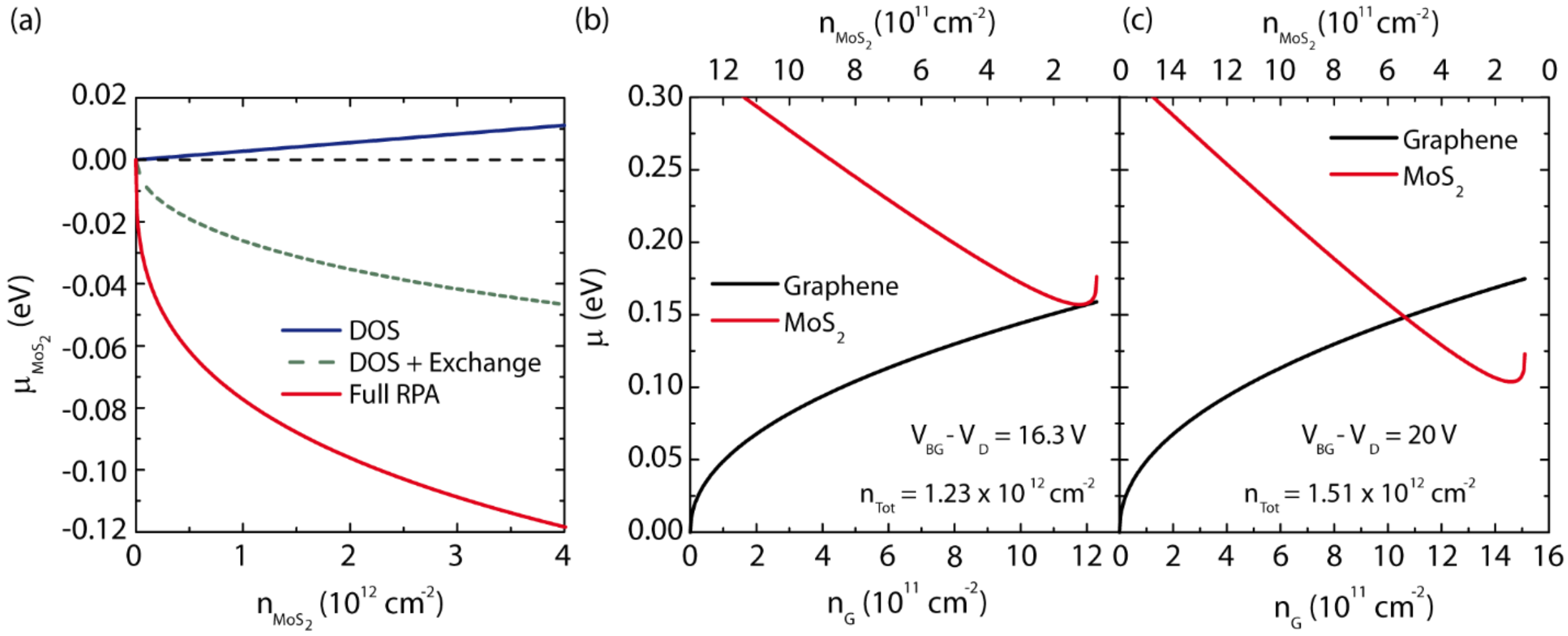

**Figure 5.** (a) $\mu_{MoS2}$ vs $n_{MoS2}$ showing the single-particle (DOS) contribution to $\mu_{MoS2}$ (blue), the DOS and exchange contribution (dashed green), and the full RPA result (red), including DOS, exchange and correlation contributions. (b-c) Graphene and $MoS_2$ chemical potentials vs. $n_G$ at fixed $n_{Tot}$, showing $\mu_G$ (black), and the RHS of Eq. (3) (red). Panel (b) corresponds to $n_{Tot} = 1.23 \times 10^{12}$ cm$^{-2}$ ($V_{BG}$ = 16.3 V), and panel (c) to $n_{Tot} = 1.5 \times 10^{12}$ cm$^{-2}$ ($V_{BG}$ = 20 V). The equilibrium condition corresponds to the intersection of the two traces, where Eq. (3) is satisfied. Panel (b-c) calculations are performed for a heterostructure with $t_{MoS2}$ = 4.2 nm, and using $\Delta E_C$ = 0.41 eV.

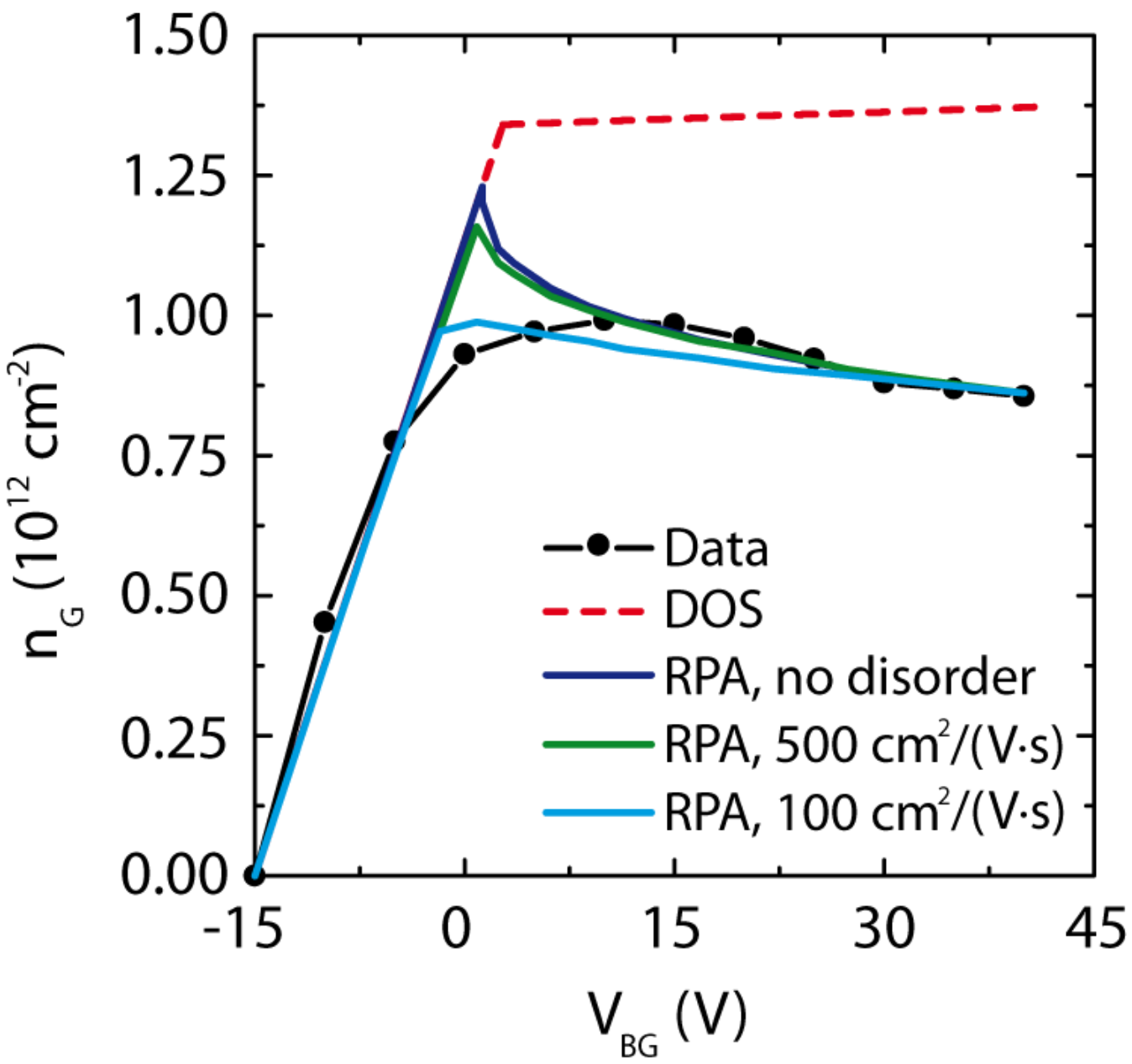

**Figure 6.** Theoretical and experimental comparison of $n_G$ vs $V_{BG}$ in a graphene-$MoS_2$ with $t_{MoS2}$ = 4.2 nm. The symbols are experimental data extracted from SdH oscillations (black dots), and the lines represent calculations performed using only the bare single-particle contribution (DOS) to $\mu_{MoS2}$ (dashed red), including interactions in RPA but not disorder (solid dark blue), RPA including disorder assuming a $MoS_2$ mobility of 500 cm$^2$/(V·s) (green) and 100 cm$^2$/(V·s) (light blue).

AUTHOR INFORMATION

**Corresponding Author**

*etutuc@mer.utexas.edu;

**Author Contributions**

The manuscript was written through contributions of all authors. All authors have given approval to the final version of the manuscript. ‡These authors contributed equally.

**Funding Sources**

This work has been supported by NRI-SWAN, ONR, Intel, and the Welch Foundation under grant TBF1473.

ACKNOWLEDGMENT

We thank K. Lee, M. Polini and J. Xue for technical discussions.